\documentclass[preprint,letterpaper,12pt]{JHEP3}
\usepackage{axodraw}
{\setlength\paperheight {11in} % \setlength\paperwidth  {8.5in}}
\setlength{\topmargin}{1in}

\newcommand{\GeV}{\,{\rm GeV}}
\newcommand{\MeV}{\,{\rm MeV}}

\newcommand{\eV}{\,{\rm eV}}
\newcommand\lsim{\mathrel{\rlap{\lower4pt\hbox{\hskip1pt$\sim$}}
    \raise1pt\hbox{$<$}}}
\newcommand\gsim{\mathrel{\rlap{\lower4pt\hbox{\hskip1pt$\sim$}}
    \raise1pt\hbox{$>$}}}

\def\bea{\begin{eqnarray}}
\def\eea{\end{eqnarray}}
\def\ba{\begin{array}}
\def\ea{\end{array}}
\def\bec{\begin{center}}
\def\ec{\end{center}}
\def\nn{\nonumber}
\def\la{\langle}
\def\ra{\rangle}

\def\ps{ SU(4)_C \times SU(2)_L \times SU(2)_R}

\def\64{\rm SO(6) \times SO(4)}

\def\f{\frac}

\def\g{\gamma}
\def\l{\lambda}
\def\f#1#2{\frac{#1}{#2}}

\setcounter{page}{1}
\preprint{OHSTPY-HEP-T-03-003\\KIAS-P03015}
\title{\huge Neutrinos in 5D SO(10) Unification}
\author{Hyung Do Kim$^{a,c}$ and Stuart Raby$^{a,b}$\\
$^a$Department of Physics, The Ohio State University,
174 W. 18th Ave., \\ Columbus, Ohio 43210, USA\\
$^b$On leave of absence, School of Natural Sciences, Institute for
Advanced Study, Princeton, NJ 08540 \\ $^c$School of Physics,Korea
Institute for Advanced Study, Seoul, 235-010, Korea
\\ E-mail: \email{hdkim,raby@mps.ohio-state.edu}}
\abstract{ We study neutrino physics in a 5D supersymmetric SO(10)
GUT.  We analyze several different choices for realizing the
See-Saw mechanism.  We find that the ``natural" scale for the
Majorana mass of right-handed neutrinos depends critically on
whether the right-handed neutrinos are located in the bulk or
localized on a brane.  In the former case, the effective Majorana
mass $M_{\rm eff}$ is ``naturally" of order the compactification
scale $M_c \sim 10^{14}$ GeV.   Note, this is the value necessary
for obtaining a light $\tau$ neutrino with $m_{\nu_\tau} \sim
10^{-2}$ eV which, within the context of hierarchical neutrino
masses, is the right order of magnitude to explain atmospheric
neutrino oscillations.   On the other-hand when the right-handed
neutrino is localized on the brane, the effective Majorana mass is
typically larger than $M_c$. Nevertheless with small parameters of
order 1/10 -- 1/30, $M_{\rm eff}\sim M_c$ can be accommodated.  We
also discuss the constraints on model building resulting from the
different scenarios for locating the right-handed neutrinos.}

\begin{document}
%%%%%%%%%%%%%%%%%%%%%%%%%%%%%%%%%%%%%%%%%%%%%%%%%%%%%%%%

%%%%%%%%%%%%%%%%%%%%%%%%%%%%%%%%%%%%%%%%%%%%%%%%%%%%%%%%
\section{Introduction}
%%%%%%%%%%%%%%%%%%%%%%%%%%%%%%%%%%%%%%%%%%%%%%%%%%%%%%%%
In a recent paper \cite{Kim:2002im},  we considered gauge coupling
unification in a five dimensional supersymmetric $SO(10)$ model
compactified on an orbifold $S^1/(Z_2 \times Z_2^{\prime})$. We
obtained an excellent prediction for gauge coupling unification
with a cutoff scale $M_* \sim 3 \times 10^{17}$ GeV and a
compactification scale $M_c \sim 1.5 \times 10^{14}$ GeV.   We
also showed that our results are mathematically equivalent to a
four dimensional supersymmetric grand unified theory [SUSY GUT]
analysis with the color triplet Higgs mass given by $M_c$
\cite{Kim:2002im}. However, unlike the 4D case, it was shown that
proton decay due to dimension 5 operators may be completely
eliminated. Therefore, in this 5D framework, the unification of
gauge couplings is elegantly explained.  At the same time all the
nice features of grand unified theories, such as charge
quantization and Yukawa unification, are maintained.
Unfortunately, the proton decay rate in these models, due to
dimension 6 operators, is sensitive to the placement of matter
multiplets in the 5th dimension, as well as to the unknown physics
above the cutoff scale.

In this paper we study neutrino masses within this same 5D
orbifold framework.  We consider the See-Saw mechanism for
neutrino masses ~\cite{grsy} and evaluate the effective
right-handed Majorana mass in different scenarios.  We find that
the ``natural" scale for the Majorana mass of right-handed
neutrinos depends critically on whether the right-handed neutrinos
are located in the bulk or localized on a brane.  In the former
case, the effective Majorana mass $M_{\rm eff}$ is ``naturally" of
order the compactification scale $M_c \sim 1.5 \times 10^{14}$
GeV.   Hence, the left-handed $\tau$ neutrino mass is of order
$m_t(m_t)^2/3 M_c \approx 0.06 \eV$\footnote{The factor of 3 in
the denominator approximately takes into account the different RG
running of the top Yukawa coupling and the effective neutrino mass
operator.} (for $m_t(m_t) \approx 165$ GeV) which is just right
(assuming hierarchical neutrino masses) to explain atmospheric
neutrino oscillations and the K2K data.

On the other-hand when the right-handed neutrino is localized on
the brane, the effective Majorana mass is typically larger than
$M_c$. Nevertheless with small parameters of order 1/10 -- 1/30,
$M_{\rm eff}\sim M_c$ can be accommodated.  We also discuss the
constraints on model building resulting from the different
scenarios for locating the right-handed neutrinos.   Of course, in
order to complete the analysis of neutrino oscillations we would
need to construct a three neutrino model. Finally, we note other
recent papers on 5D $SO(10)$ SUSY GUTs within a similar orbifold
framework \cite{Dermisek:2001hp} - \cite{Kitano:2003cn}. Also an
alternate mechanism for obtaining small neutrino masses in 5D GUTs
is given in \cite{Hebecker:2002re,Shafi:2003ie}.

%%%%%%%%%%%%%%%%%%%%%%%%%%%%%%%%%%%%%%%%%%%%%%%%%%%%%%%%%
\section{See-Saw and Double See-Saw Mechanism in 4D}
%%%%%%%%%%%%%%%%%%%%%%%%%%%%%%%%%%%%%%%%%%%%%%%%%%%%%%%%%

%%%%%%%%%%%%%%%%%%%%%%%%%%%%%%%%%%%%%%%%%%%%%%%%%%%%%%%%%
\subsection{See-Saw Mechanism}
%%%%%%%%%%%%%%%%%%%%%%%%%%%%%%%%%%%%%%%%%%%%%%%%%%%%%%%%%

Quarks and charged leptons have mass ranging from ${\cal
O}(\frac{1}{2} \ \MeV$ to $175 \ \GeV)$, which is determined
theoretically in terms of a dimensionless Yukawa coupling, ${\cal
O}(10^{-6}$ to 1), times a Higgs vev $v \approx 246 \GeV$. The
heaviest neutrino mass $\lsim \eV$ requires a neutrino - Higgs
Yukawa coupling ${\cal O} (10^{-12})$ which is unnaturally small.
The GRSY See-Saw mechanism, on the other hand, provides a natural
explanation for such small neutrino masses with order one Yukawa
couplings~\cite{grsy}. Since the right-handed neutrino $\nu_R$
(necessary for a Dirac neutrino mass) has zero charge under all SM
gauge interactions, it can have a Majorana mass $M$ much larger
than the electroweak scale.

The simple See-Saw mass matrix for the neutrino is given by
\bea  M_{\nu} \left(%
\begin{array}{c}
  \nu \\
  \nu^c \\
\end{array}%
\right) & = &
 \left(%
\begin{array}{cc}
  0 & m_D \\
  m_D & M \\
\end{array}%
\right)
\left(%
\begin{array}{c}
  \nu \\
  \nu^c \\
\end{array}%
\right) \eea where $\nu^c \equiv \nu_R^*$.  The natural scale for
$M$ (in a 4D GUT) is of order the GUT or cutoff scale with $M \gg
m_D$. Hence the two eigenvalues are very different with
\bea m_1 & \simeq &  \f{m_D^2}{M}, \\
m_2 & \simeq & M, \nn \\
m_1 & \ll & m_D \ll m_2. \nn \eea  Thus the largeness of $M$
naturally explains the smallness of the left-handed neutrino mass.
The problem is that in order to explain the low energy neutrino
data, $M$ must be at least two orders of magnitude smaller than
the SUSY GUT scale.

%%%%%%%%%%%%%%%%%%%%%%%%%%%%%%%%%%%%%%%%%%%%%%%%%%%%%%%%%
\subsection{Double See-Saw Mechanism}
%%%%%%%%%%%%%%%%%%%%%%%%%%%%%%%%%%%%%%%%%%%%%%%%%%%%%%%%%

In SO(10) GUTs, all quarks and leptons in one generation belong to
a single spinor representation $\bf 16 \;\; [= 10 + \bar{5} +1
\;\;$ of SU(5)]. Unlike SU(5) where the minimal matter content
includes only $\bf 10$s and $\bf \bar{5}$s with no right-handed
neutrinos, SO(10) has a right-handed neutrino, neutral under the
standard model gauge group, which is contained in the (minimal)
${\bf 16}$ representation. Hence the simple See-Saw mechanism is
not applicable in an SO(10) model, since the right-handed neutrino
is not a singlet under SO(10); thus making it impossible to have a
renormalizable Majorana mass.

For the See-Saw mechanism to work in an SO(10) model, we need to
give the right-handed neutrino a Majorana mass of order the GUT
scale.   There are two methods for obtaining this:
\begin{itemize} \item using a higher dimension operator
\begin{equation} \f{\overline{16} \ 16 \ \overline{16} \  16}{M_*}, \;\; {\rm
or} \end{equation} \item adding an $SO(10)$ singlet $N$ and the
renormalizable interactions \begin{equation} \lambda_N \
\overline{16} \ 16 \ N + \f{1}{2} \ M_2 \ N \ N \end{equation}
\end{itemize}  where $M_*$ is the cutoff scale of the theory.
In both cases we assume a non-zero vacuum expectation value [vev]
$$\lambda_N \ \langle {\bf \overline{16}} \rangle  = M_1 \neq 0$$ in the right-handed neutrino
direction.   In the first case, the product of fields has, in
general, several inequivalent SO(10) invariant combinations. One
particularly simple combination in the first case, with the first
two fields combined to make an SO(10) singlet, can be obtained as
an effective interaction after integrating out $N$ in the second
case. Note that the effective See-Saw mass $M_{\rm eff}$, in
either case, can be lower than the cutoff scale of the theory.  In
this paper we consider the latter method.

We now include the usual, electroweak scale, Dirac mass coming
from the Yukawa term $ \lambda \ 16_3 \ 10_H \ 16_3$ (with 3
denoting the third generation).  After electroweak symmetry
breaking we have \bea W & = & m_D \ \nu \ \nu^c \eea (with $m_D =
\l \f{v}{\sqrt{2}} \sin\beta$) and we obtain a $3 \times 3$
neutrino mass matrix, rather than a $2 \times 2$
matrix,\footnote{This is similar to the double see-saw mechanism
suggested by Mohapatra and Valle~\cite{doubleseesaw}.} given by
\bea  {\cal M} \left(%
\begin{array}{c}
  \nu \\
  \nu^c \\
  N \\
\end{array}%
\right) & = & \left(%
\begin{array}{ccc}
  0 & m_D & 0 \\
  m_D & 0 & M_1 \\
  0 & M_1 & M_2 \\
\end{array}%
\right)
\left(%
\begin{array}{c}
  \nu \\
  \nu^c \\
  N \\
\end{array}%
\right). \eea  Note, $|{\rm Det} {\cal M}| =  m_D^2 M_2$, Tr
${\cal M} = M_2$ and $m_D \ll M_1, M_2$.  Hence the effective
See-Saw scale $M_{\rm eff}$ in this case may be obtained by
evaluating the inverse of the heavy $2 \times 2$ mass matrix  \bea
\left(
\begin{array}{cc}
   0 & M_1 \\
   M_1 & M_2 \\
\end{array} \right) \eea
in the $\nu^c$ direction.  We find $M_{\rm eff} = M_1^2/M_2$.
Note, the result is independent of the mass ordering, i.e. $M_1
\ll M_2$, $M_1 \gg M_2$ or $M_1 \approx M_2$.  Finally, we obtain
the light neutrino mass given by \bea m_{\nu} & \simeq & \f{m_D^2
M_2}{M_1^2}, \eea irrespective of the ratio $M_1/M_2$, as long as
$m_D \ll M_1, M_2$. The question now is how to obtain an effective
See-Saw scale $M_{\rm eff} = M_1^2/M_2 \sim 10^{14} \ \GeV$.

%%%%%%%%%%%%%%%%%%%%%%%%%%%%%%%%%%%%%%%%%%%%%%%%%%%%%%%%%
\section{Setup -- SO(10) on ${\cal M}_4 \times S^1/(Z_2 \times Z_2^\prime)$}
%%%%%%%%%%%%%%%%%%%%%%%%%%%%%%%%%%%%%%%%%%%%%%%%%%%%%%%%%

We consider a five dimensional supersymmetric $SO(10)$ GUT
compactified on an \\ $S^1/(Z_2 \times Z_2^{\prime})$ orbifold.
One orbifolding, $Z_2$, reduces $N=2$ supersymmetry to $N=1$, and
the other, $Z_2^{\prime}$, breaks $SO(10)$ to the Pati-Salam [PS]
gauge group $\ps$.  The space time in the 5th direction is the
line segment $y: [0, \pi R/2]$ with an $SO(10)$ symmetry in the
bulk and on the brane at $y = 0$, but only $\ps$ symmetry on the
brane at $y = \pi R/2$. We call these two fixed points
respectively, ``SO(10)" and ``Pati-Salam" branes. The brane
describes a $3+1$ dimensional spacetime and the gauge group is the
maximal gauge symmetry surviving on each brane. Further breaking
to the standard model gauge group is accomplished via the Higgs
mechanism on the PS brane \cite{Kim:2002im}.

We are interested in obtaining the Majorana mass scale for
right-handed neutrino masses.  The natural candidates for this
scale are the cutoff scale  $M_* \sim 3 \times 10^{17}$ GeV and
the compactification scale $M_c \sim 1.5 \times 10^{14}$ GeV.  Let
us consider just one family of quarks and leptons (including the
top, bottom and tau) which we consider locating either on the PS
brane or in the bulk. Quarks and leptons of one SM family are
contained in two irreducible representations given by the
left-handed Weyl spinors, \bea \psi
\equiv & \{ \; Q\; \;, \; \; \; L \;\} \;\; & \subset \;\; {\bf (4, 2, 1)}, \nn \\
\psi^c \equiv & \{ \left( \begin{array}{c} t^c \\ b^c
\end{array}\right) , \; \left( \begin{array}{c} \nu_\tau^c \\
\tau^c \end{array}\right)  \} \;\; &\subset \;\; {\bf (\bar 4,
1, \bar 2)}, \nn \eea where $Q = \left( \begin{array}{c} t \\ b
\end{array}\right)$ and  $L = \left( \begin{array}{c} \nu_\tau \\
\tau \end{array}\right)$ are left-handed electroweak doublets.
There are also brane fields $\chi^c$ in $\bf (\bar 4,1,\bar 2)$
and $\bar \chi^c$ in $\bf (4,1,2)$ which break Pati-Salam down to
the Standard Model gauge group by getting vacuum expectation
values [vev]s along the right-handed neutrino direction.  Note, we
assume that the vev of $\chi^c$ and $\bar \chi^c$ and all the
parameters appearing in the model are of order one in an
appropriate unit set by the cutoff scale $M_*$.

With regards to gauge coupling unification, whether one has bulk
matter fields and/or brane localized matter fields does not make a
difference as long as one has complete SO(10) multiplets at every
KK level. However, once there are additional hypermultiplets in
the bulk, the 5D gauge theory rapidly approaches a strong coupling
regime. In our previous work \cite{Kim:2002im} it was shown that
the unification of gauge couplings is only achieved with a
moderate hierarchy between the unification scale $M_*$ and the
compactification scale $M_c = \f{1}{R}$ with $M_* R \sim 10^3$.
This moderately large ratio of $M_* / M_c$ is self-consistent only
if a non-trivial fixed point exists in the 5D super-Yang-Mills
theory. The condition for such a strongly coupled fixed point in
5D is satisfied only when the number of SO(10) hypermultiplets in
the bulk with $n_{10}$ in vector and $n_{16}$ in spinor
representations satisfies $n_{10} \le 6$ and $n_{16} \le
2$~\cite{Seiberg:1996bd}. Therefore, in order to maintain
self-consistency, we assume that at most one generation of matter
fields, requiring two ${\bf 16}$ hypermultiplets of SO(10), are in
the bulk. This restriction does not, however, apply to SO(10)
singlets.  We introduce the SO(10) singlet, bulk fields $N$ and
$N^c$ which form a hypermultiplet in 5D.

There are two choices for the orbifold parity of $N$ and
$N^c$.{\footnote{We use $N$ and $N^c$ for both superfields and
their fermionic components. Here they are superfields.}}
\begin{itemize}
\item $N_{(+,+)}$ and $N^c_{(-,-)}$

$N^c$ vanishes on both SO(10) and Pati-Salam fixed points.

\item $N_{(+,-)}$ and $N^c_{(-,+)}$

$N$ vanishes on Pati-Salam fixed point and $N^c$ vanishes on
SO(10) fixed point. The Fourier expansions of $N$ and $N^c$ are
given by \bea N & = & \frac{1}{\sqrt{\pi R}} \sum_{n=0}^{\infty}
N_n \cos
(\frac{(2n+1)y}{R}), \label{eq:nkk} \\
N^c & = & \frac{1}{\sqrt{\pi R}} \sum_{n=0}^{\infty} N^c_n \sin
(\frac{(2n+1)y}{R}). \nn \eea
\end{itemize}
From now on, we consider the second case in which the zero modes
of $N$ and $N^c$ are already projected out by the orbifolding.

%%%%%%%%%%%%%%%%%%%%%%%%%%%%%%%%%%%%%%%%%%%%%%%%%%%%%%%%%
\section{Different Mass Terms in 5D}
%%%%%%%%%%%%%%%%%%%%%%%%%%%%%%%%%%%%%%%%%%%%%%%%%%%%%%%%%

For $\nu$, $\nu^c$, $N$ and $N^c$, we have various mass terms
either in the bulk or on the brane. For $N$ and $N^c$, we can have
both bulk mass terms and brane mass terms. Furthermore, they have
5D kinetic terms with $\partial_5$ which behave as 4D mass terms.
The fields $\nu$ and $\nu^c$ can have brane localized mass terms
by themselves or with $N$ and $N^c$. Since some fields reside only
on the Pati-Salam brane and we have 5D $N=1$ supersymmetry in the
bulk, not all mass terms are allowed.  Let us thus discuss the
allowed neutrino mass terms in a 5D $N=1$ supersymmetric theory
with orbifold fixed points.

%%%%%%%%%%%%%%%%%%%%%%%%%%%%%%%%%%%%%%%%%%%%%%%%%%%%%%%%%
\subsection*{Dirac Mass}
%%%%%%%%%%%%%%%%%%%%%%%%%%%%%%%%%%%%%%%%%%%%%%%%%%%%%%%%%
\begin{itemize}

\item{\em Bulk Kaluza-Klein mass for $N$ and $N^c$}

There are Kaluza-Klein mass terms for $N,\ N^c$ which come from a
5D kinetic term,
\bea \int dy W & = & \int dy \ N \ \partial_5 N^c  \label{eq:bulkdirac} \\
& \approx & \sum_{n=0}^{N_*} \ (2n+1)M_c \ N_n \ N^c_n \nn \eea

Note, consistent with the cutoff scale $M_*$, we truncate KK modes
at $N_*$ satisfying $2 N_* M_c \simeq M_*$.

\item{\em Kink mass for $N$ and $N^c$}

We can write a Dirac mass term for $N$ and $N^c$ in the bulk, \bea
W & = & M_{\rm kink} \ N \ N^c, \eea where $M_{\rm kink}$ must be
odd under $Z_2 \times Z_2^{\prime}$; hence a kink mass. Depending
on the sign of $M_{\rm kink}$, there can appear two new light
modes localized at opposite fixed points. In this paper we do not
consider this possibility and set $M_{\rm kink} = 0$

\item{\em SM Yukawa couplings}

We can write down the usual Yukawa couplings on the Pati-Salam
brane. \bea  W & = &  \lambda \ \psi \ {\cal H} \ \psi^c \ \delta
(y - \frac{\pi R}{2}) \label{eq:so10mass} \eea

When ${\cal H} \equiv \bf (1,\bar 2,2)$ acquires a vev $v \approx
246$ GeV at the electroweak scale with $F_{\cal H} =0$, we obtain
the usual Dirac mass terms between the left-handed $\nu$ and the
right-handed $\nu^{c *}$, \bea W & = & m \ \nu \ \nu^c \ \delta(y-
\frac{\pi R}{2}) \label{eq:nudirac} \eea with $m = \l
\f{v}{\sqrt{2}} \sin\beta$.

\item{\em Pati-Salam Yukawa couplings}

Both bulk and brane fields can have interactions on the brane. The
PS gauge symmetry and the orbifold parity restrict the possible
interactions.

We consider two cases: i) $\psi^c_3$ in the bulk and ii)
$\psi^c_3$ on the brane.

\begin{itemize}
\item{\em $\psi^c_3$ in the bulk}

If $\psi^c_3$ is in the bulk, the relevant interaction is given by
\bea W & = & \f{6 \pi^2}{M_*} \l_D \ \bar \chi^c \ \psi^c N^c \
\delta (y - \frac{\pi R}{2}). \label{eq:psdiracbulk} \eea The
coefficient is determined such that $\l_D=1$ corresponds to the
result from naive dimensional analysis with a strong coupling
assumption.\footnote{You can easily obtain this result by having
an overall factor of $1/(16 \pi^2)$ and then multiply by $4\pi$
for each brane field and $(24 \pi^3)^{1/2}$ for each bulk field in
the expression~\cite{nda}.} We consider $\l_D=1$ as a ``natural"
value of the coupling. When $\bar \chi^c$ develops a vev at the
cutoff scale with $F_{\bar \chi^c} =0$ and $\la \bar \chi^c \ra
\simeq \f{M_*}{4\pi}$, we get \bea W & = & \f{3\pi}{2} \ \l_D \
\nu^c \ N^c \ \delta (y - \frac{\pi R}{2}). \eea We then decompose
it into KK modes, \bea \int dy W & = & \sqrt{\f{3}{2
2^{\delta_{n,0}}}} \l_D M_c \ \sum_{n,m=0}^{N_*} \
(-1)^{n+m} \nu^c_n \ N^c_m  \\
& = & \sqrt{\f{2}{2^{\delta_{n,0}}}} \sum_{n,m=0}^{N_*} \
(-1)^{n+m} M_D \ \nu^c_n \ N^c_m. \nn \eea Finally the Dirac
neutrino mass for $\nu^c$ and $N_n$ is given by \bea M_D & = &
\sqrt{\f{3}{4}} \l_D M_c, \label{eq:mdbulk} \eea

For $\l_D=1$, we have the ``natural" value for $M_D =
\sqrt{\f{3}{4}} M_c \simeq M_c$, using, as stated above, naive
dimensional analysis.

\item{\em $\psi^c_3$ on the PS brane}

If $\psi^c_3$ is on the PS brane, the interaction is\footnote{Note
$\l_D$ defined in Eqn. \ref{eq:psdiracbrane} is not the same
quantity defined in Eqn. \ref{eq:psdiracbulk}, although we use the
same notation.  This should not cause any confusion, since we
never use both at the same time.} \bea W & = &
\sqrt{\f{24\pi^3}{M_*}} \l_D \ \bar \chi^c \ \psi^c N^c \ \delta
(y - \frac{\pi R}{2}). \label{eq:psdiracbrane} \eea With $\la \bar
\chi^c \ra \simeq \f{M_*}{4\pi}$, we get \bea W & = &
\sqrt{\f{3\pi M_*}{2}} \ \l_D \ \nu^c \ N^c  \ \delta (y -
\frac{\pi R}{2}). \eea  KK decomposition gives \bea \int dy W & =
& \sqrt{\f{3M_* M_c}{2}} \l_D \ \sum_{n=0}^{N_*} \
(-1)^n \nu^c \ N^c_n  \\
& = & \sum_{n=0}^{N_*} \ (-1)^{n} M_D \ \nu^c \ N^c_n. \nn \eea
The Dirac mass $M_D$ for $\nu^c$ and $N_n$ is then given by \bea
M_D & = & \sqrt{\f{3M_*}{2M_c}} \l_D M_c. \label{eq:mdbrane} \eea

In this case, for $\l_D =1$, we have $M_D = \sqrt{\f{3M_*}{2M_c}}
M_c \sim \sqrt{\f{M_*}{M_c}} M_c \sim 30 \ M_c$ for $M_*/M_c \sim
10^3$.

\end{itemize}

Note, although the term \bea W^\prime & = & C \ \l_D^{\prime} \
\bar \chi^c \
\partial_5 N \ \psi^c \ \delta (y - \frac{\pi R}{2})  \eea is also
possible, with $C = \f{6\pi^2}{M_*^2}$ for $\psi^c$ in the bulk
and $C = \sqrt{\f{24\pi^3}{M_*^3}}$ for $\psi^c$ on the PS brane,
it is subleading compared to the term without the derivative
$\partial_5$ and is suppressed by $M_c/M_*$.  We do not consider
this term further in the paper.

\end{itemize}

\subsection*{Majorana Mass}

\begin{itemize}

\item{\em Majorana Mass in the bulk}

A Majorana mass is allowed in the bulk for gauge singlet fields,
\bea W & = & \f{1}{2} M_N ( -N N + N^c N^c ). \label{eq:mnbulk1}
\eea A 5D Majorana mass has a relative minus sign when it is
expressed in terms of two 4D Weyl spinors.\footnote{It is possible
to redefine $\tilde{N} = -i N$ such that the Majorana mass does
not have a relative minus sign, $W = \f{1}{2} M_N ( \tilde{N}
\tilde{N} + N^c N^c )$. In this case the Dirac (KK) mass term
becomes $ W = i \tilde{N} \partial_5 N^c$. In this paper, we use
the Dirac (KK) mass term $W = N \partial_5 N^c$.} We consider all
possible values of $M_N \leq M_*$.  (See the appendix for a
detailed discussion of Majorana masses in 5D.)

\item{\em Majorana Mass on the brane}

It is possible to have a brane Majorana mass, \bea W & = &
\f{1}{2} (a_1 N N + \f{d_1}{M_*^2} \partial_5 N^c \partial_5 N^c )
\delta (y) \nn \\ &&+ \f{1}{2} (a_2 N^c N^c + \f{d_2}{M_*^2}
\partial_5 N
\partial_5 N ) \delta (y-\f{\pi R}{2}). \label{eq:mnbrane1} \eea

However, brane Majorana mass terms are volume suppressed compared
to the bulk Majorana mass by a factor $M_c/M_*$. In addition,
terms with $\partial_5$ are suppressed by extra powers of
$M_c/M_*$. Therefore, we set the brane Majorana mass terms to zero
in our analysis when considering a bulk Majorana mass. However,
when there is no bulk Majorana mass, the brane Majorana mass plays
an important role.

\end{itemize}

Let us now summarize the mass terms in our setup. We have a bulk
Dirac mass for $N_{(+,-)}$ and $N^c_{(-,+)}$ coming from the 5D
kinetic term (Eqn. \ref{eq:bulkdirac}) and also a bulk (Eqn.
\ref{eq:mnbulk1}) or brane (Eqn. \ref{eq:mnbrane1}) Majorana mass.
We also have a Dirac mass coupling $N$ to $\nu^c$ on the PS brane.
This takes different values depending on whether $\psi_3^c$ is
located in the bulk (Eqns. \ref{eq:psdiracbulk} - \ref{eq:mdbulk})
or for $\psi_3^c$ on the brane (Eqns. \ref{eq:psdiracbrane} -
\ref{eq:mdbrane}). As a consequence of these mass terms, we obtain
an effective Majorana mass for $\nu^c$.  The fields $\nu, \ \nu^c$
then obtain a Dirac mass, at the electroweak scale via the Higgs
doublet vev (Eqns. \ref{eq:so10mass} - \ref{eq:nudirac}).

%%%%%%%%%%%%%%%%%%%%%%%%%%%%%%%%%%%%%%%%%%%%%%%%%%%%%%%%%
\section{Neutrino Mass Matrix}
%%%%%%%%%%%%%%%%%%%%%%%%%%%%%%%%%%%%%%%%%%%%%%%%%%%%%%%%%

We now calculate the eigenvalues of the neutrino mass matrix. The
aim of this paper is to evaluate the left-handed tau neutrino mass
using the See-Saw mechanism. We do this in two steps. Step 1:
Calculate the effective Majorana mass for the right-handed tau
neutrino $\nu^c_\tau$. Since the electroweak Dirac neutrino mass
($m$) is extremely small compared to all other scales ($M_*$ or
$M_c$), we first deal with the mass matrix for $\nu^c$, $N$ and
$N^c$. Step 2: We then calculate the mass of $\nu_\tau$. We use
techniques similar to those discussed in
\cite{Dienes:1998sb,Arkani-Hamed:1998vp}. Note, in the following
the left- (right)-handed tau neutrino are simply denoted by $\nu \
(\nu^c)$.

There are two possibilities for the Majorana mass of $N$ and/or
$N^c$. It can come from either a bulk Majorana mass or from a
brane Majorana mass. We consider both possibilities here.   We
also consider the two choices of locating $\nu^c$ either in the
bulk or on the PS brane.   We show that the final result is
independent of whether we use a bulk or brane mass for $N$ and/or
$N^c$.  However the result depends critically on whether the field
$\nu^c$ is located in the bulk or on the PS brane.  In the former
case we show that $M_{\rm eff} \approx M_c$ ``naturally?"   In the
latter case, we describe how to obtain this result with minimal
tuning of parameters.

%%%%%%%%%%%%%%%%%%%%%%%%%%%%%%%%%%%%%%%%%%%%%%%%%%%%%%%%%%%%%%%%%
\subsection{Bulk Majorana Mass for $N$ and $N^c$}
%%%%%%%%%%%%%%%%%%%%%%%%%%%%%%%%%%%%%%%%%%%%%%%%%%%%%%%%%%%%%%%%%
The most general superpotential has the mass terms as \bea W & = &
N \partial_5 N^c + \f{1}{2} M_N (- N N + N^c N^c) + C \l_D \nu^c
N^c \delta (y-\f{\pi R}{2}), \eea which is decomposed into KK
modes as \bea W & = & \sum_{n=0}^{N_*} \left[ (2n+1) M_c N_n N^c_n
+ \f{1}{2} M_N (- N_n N_n + N^c_n N^c_n) + (-1)^n M_D \nu^c N^c_n
\right]. \eea   Note, the constant $C = \f{3 \pi}{2} \;\;
(\sqrt{\f{3 \pi M_*}{2}})$ for $\nu^c$ in the bulk (on the PS
brane).

%%%%%%%%%%%%%%%%%%%%%%%%%%%%%%%%%%%%%%%%%%%%%%%%%%%%%%%%%%%%%
\subsubsection{Brane localized $\nu^c$ case}
%%%%%%%%%%%%%%%%%%%%%%%%%%%%%%%%%%%%%%%%%%%%%%%%%%%%%%%%%%%%%%

The above formula applies to both the bulk $\nu^c$ and the brane
localized $\nu^c$.   In the following discussion we first consider
the case of a brane localized $\nu^c$,  however our discussion
below applies equally well to the zero mode of a bulk field with
the identification $\nu^c \equiv \nu^c_0$.  Later we will add the
tower of KK modes for the bulk $\nu^c$ case.

We now define \bea {\cal N} & = & (\nu^c, N_0, N^c_0, N_1, N^c_1,
\cdots, N_n, N^c_n, \cdots). \eea Then \bea W & = & \f{1}{2} \
{\cal N}^T \ \hat{\cal M} \ {\cal N} \eea with \bea \hat{\cal M} &
= & \left( \ba{ccccccccc} 0 & 0 & M_D & 0 & -M_D &
\cdots & 0 & (-1)^n M_D & \cdots \\
0 & -M_N & M_c & 0 & 0 & \cdots & 0 & 0 & \cdots
\\ M_D & M_c & M_N & 0 & 0 & \cdots & 0 & 0 & \cdots \\
0 & 0 & 0 & -M_N & 3M_c & \cdots & 0 & 0 & \cdots \\
-M_D & 0 & 0 & 3M_c & M_N & \cdots & 0 & 0 & \cdots \\
\cdots & \cdots & \cdots & \cdots & \cdots & \cdots & \cdots &
\cdots & \cdots \\
0 & 0 & 0 & 0 & 0 & \cdots & -M_N & (2n+1)M_c & \cdots \\
(-1)^n M_D & 0 & 0 & 0 & 0 & \cdots & (2n+1)M_c & M_N & \cdots \\
\cdots & \cdots & \cdots & \cdots & \cdots & \cdots & \cdots &
\cdots & \cdots
 \ea \right). \eea

The effective right-handed neutrino mass is obtained by finding
$({\hat{\cal M}}^{-1})_{\nu^c \nu^c}$. Let ${\cal M}$ be the
matrix without $\nu^c$ which has one less column and row compared
to $\hat{\cal M}$.

\bea ({\hat{\cal M}}^{-1})_{\nu^c \nu^c} & = & \frac{{\rm Det}
{\cal M}}{{\rm Det} \hat{\cal M}}. \eea

\bea {\rm Det} {\cal M} & = & \prod_{n=0}^{N_*}
(-M_N^2-((2n+1)M_c)^2). \eea Using the formulae in the appendix we
obtain \bea {\rm Det} \hat{\cal M} & = & M_D^2 M_N
\sum_{n=0}^{N_*}
\f{1}{M_N^2+((2n+1)M_c)^2} {\rm Det} {\cal M} \nn \\
& \approx &  \f{\pi M_D^2}{4M_c} \tanh (\f{\pi M_N}{2M_c}) {\rm
Det} {\cal M}. \eea  Then \bea M_{\rm eff} & = & \f{1}{({\hat{\cal
M}}^{-1})_{\nu^c \nu^c}} = \f{\pi M_D^2}{4M_c} \tanh (\f{\pi
M_N}{2M_c}) \label{eq:meff1} \eea is the effective Majorana mass
for the right-handed neutrino. Note, the limit $N_* \rightarrow
\infty$ is finite and the equality is obtained in this limit.

We now consider two possible limits for the bulk Majorana mass
$M_N$, i.e. [Case (1)] $M_c \le M_N \le M_*$ and [Case (2)] $M_N
\ll M_c$.  In case (1), the answer is insensitive to the size of
$M_N$. We obtain the effective right-handed neutrino mass \bea
M_{\rm eff} & = & \f{\pi M_D^2}{4M_c}. \eea In case (2), we have
\bea M_{\rm eff} & = &  \f{\pi^2 M_D^2 M_N}{8M_c^2}.
\label{eq:meffmnsmall} \eea In order to further determine the size
of $M_{\rm eff}$ we now consider the natural size of the Dirac
mass $M_D$.   It is in the calculation of $M_D$ that the location
of $\nu^c$ becomes relevant.

We first consider case (1).  When $\nu^c$ is on the PS brane, the
natural size of $M_D = \sqrt{\f{3M_*}{2M_c}} \l_D M_c$ (Eqn.
\ref{eq:mdbrane}) with $\l_D$ ``naturally" {\cal O}(1).  Hence the
``natural" value for $M_{\rm eff} \sim M_*$.  In order to obtain
the correct size for the right-handed Majorana mass we need $\l_D
\sim \sqrt{\f{M_c}{M_*}} \sim \f{1}{30}$. This is not a
particularly onerous amount of fine-tuning. However, this may be
achieved ``naturally" by adding a brane localized SO(10) singlet
field $S$ and replacing the Dirac mass term (Eqn.
\ref{eq:psdiracbrane}) by the mass term \bea W = & \sqrt{\f{24
\pi^3}{M_*}} \f{S}{M_*} \bar \chi^c \psi^c N^c \ \delta(y - \f{\pi
R}{2}). & \eea  Note, the leading term, $W = \bar \chi^c \psi^c
N^c$ can be forbidden by a U(1) symmetry under which $S$ and
$\psi^c$ carry charges $1,-1$ respectively. In this case, the
fundamental Yukawa coupling $\l_D$ is replaced by the ratio $\f{4
\pi \la S \ra}{M_*}$.  The suppression factor {\cal
O}($\f{1}{30}$) is now obtained ``naturally" by a U(1) symmetry
breaking vev for $\f{4 \pi \la S \ra}{M_*}$, i.e. one order of
magnitude less than its ``natural" value.

In case (2), where \bea M_{\rm eff} & = & \f{\pi^2 M_D^2
M_N}{8M_c^2} = \f{3\pi^2}{16} \l_D^2 \f{M_*}{M_c} M_N,  \eea we
need $\l_D^2 \ M_N \approx \f{M_c}{M_*} M_c \approx 10^{-3} M_c$.
This can be obtained with either a small Majorana mass $M_N
\approx 10^{-3} M_c$, or a small Yukawa coupling $\l_D \sim
\f{1}{30}$ or some linear combination thereof such as $\l_D \sim
\f{M_N}{M_c} \sim \f{1}{10}$. {\em In any case the desired value
for $M_{\rm eff} \approx M_c$ can be accommodated.}

%%%%%%%%%%%%%%%%%%%%%%%%%%%%%%%%%%%%%%%%%%%%%%%%%%%%%%%%%%%%%%%%%
\subsubsection{Bulk $\nu^c$ case : Inclusion of $\nu^c$ KK modes}
%%%%%%%%%%%%%%%%%%%%%%%%%%%%%%%%%%%%%%%%%%%%%%%%%%%%%%%%%%%%%%%%%
We now show that in the case of $\nu^c$ in the bulk, although the
formula for $M_{\rm eff}$ (Eqn. \ref{eq:meff1}) is unchanged, the
``natural" value for the Dirac mass $M_D \approx M_c$.

In the previous case we assumed $\nu^c$ is a brane field without
KK modes.  Now, when $\nu^c$ (i.e. $\psi^c$) is in the bulk, the
mass matrix becomes much larger, but the analysis remains the
same.  The mass term is given by \bea W & = & \f{1}{2} \ {\cal
N}^T \ \hat{\cal M} \ {\cal N}, \eea with an extended definition
of $\cal N$, \bea {\cal N} & = & (\nu^c_0, N_0, N^c_0, N_1, N^c_1,
\cdots, \nu^c_1, \bar{\nu^c}_1, \nu^c_2, \bar{\nu^c}_2, \cdots),
\eea and the mass matrix is \bea \hat{\cal M} & = & \left(
\ba{ccccccccccc} 0 & 0 & M_D & 0 & -M_D &
\cdots & 0 & 0 & 0 & 0& \cdots \\
0 & -M_N & M_c & 0 & 0 & \cdots & 0 & 0 & 0 & 0 & \cdots
\\ M_D & M_c & M_N & 0 & 0 & \cdots & -\sqrt{2}M_D & 0 & \sqrt{2}M_D & 0 & \cdots \\
0 & 0 & 0 & -M_N & 3M_c & \cdots &  0 & 0 & 0 & 0 & \cdots \\
-M_D & 0 & 0 & 3M_c & M_N & \cdots & \sqrt{2}M_D & 0 & -\sqrt{2}M_D & 0 & \cdots \\
\cdots & \cdots & \cdots & \cdots & \cdots & \cdots & \cdots &
\cdots & \cdots & \cdots & \cdots \\
0 & 0 & -\sqrt{2}M_D & 0 & \sqrt{2}M_D & \cdots & 0 & 2M_c & 0 & 0 & \cdots \\
0 & 0 & 0 & 0 & 0 & \cdots & 2M_c & 0 & 0 & 0 & \cdots \\
0 & 0 & \sqrt{2}M_D & 0 & -\sqrt{2}M_D & \cdots & 0 & 0 & 0 & 4M_c & \cdots \\
0 & 0 & 0 & 0 & 0 & \cdots & 0 & 0 & 4M_c & 0 & \cdots \\
\cdots & \cdots & \cdots & \cdots & \cdots & \cdots & \cdots &
\cdots & \cdots & \cdots & \cdots
 \ea \right) \nonumber . \\ & &  \eea

The effective right-handed neutrino mass is obtained by finding
$({\hat{\cal M}}^{-1})_{\nu^c_n \nu^c_n}$ for $n=0,1,2,\cdots$. We
find  \bea M_{\rm \nu^c_0 \nu^c_0} & = & \f{1}{({\hat{\cal
M}}^{-1})_{\nu^c_0 \nu^c_0}} = \f{\pi M_D^2}{4M_c} \tanh
(\f{\pi M_N}{2M_c}) , \label{eq:meff} \\
({\hat{\cal M}}^{-1})_{\nu^c_n \nu^c_n} & = & 0 \ \ \
(n=1,2,\cdots). \eea   We note two interesting facts about this
result. Firstly, this result is identical to the effective
right-handed neutrino Majorana mass found for the previous case
with $\nu^c$ localized on the brane (Eqn. \ref{eq:meff1}).
Secondly, the effective Majorana mass is generated only for the
zero mode of $\nu^c$ and is independent of KK modes of $\nu^c$.
You can easily check this using the above mass matrix by adding or
subtracting $\sqrt{2} \times \nu^c_0$ column (row) to the
$\nu^c_n$ column (row) which makes the mass matrix block diagonal
form for $\nu^c_n$ for $n \ge 1$. This operation does not change
the determinant.   We find \bea {\rm Det} \hat{\cal M} & = & {\rm
Det} \left( \ba{ccccccccccc} 0 & 0 & M_D & 0 & -M_D &
\cdots & 0 & 0 & 0 & 0& \cdots \\
0 & -M_N & M_c & 0 & 0 & \cdots & 0 & 0 & 0 & 0 & \cdots
\\ M_D & M_c & M_N & 0 & 0 & \cdots & 0 & 0 & 0 & 0 & \cdots \\
0 & 0 & 0 & -M_N & 3M_c & \cdots &  0 & 0 & 0 & 0 & \cdots \\
-M_D & 0 & 0 & 3M_c & M_N & \cdots & 0 & 0 & 0 & 0 & \cdots \\
\cdots & \cdots & \cdots & \cdots & \cdots & \cdots & \cdots &
\cdots & \cdots & \cdots & \cdots \\
0 & 0 & 0 & 0 & 0 & \cdots & 0 & 2M_c & 0 & 0 & \cdots \\
0 & 0 & 0 & 0 & 0 & \cdots & 2M_c & 0 & 0 & 0 & \cdots \\
0 & 0 & 0 & 0 & 0 & \cdots & 0 & 0 & 0 & 4M_c & \cdots \\
0 & 0 & 0 & 0 & 0 & \cdots & 0 & 0 & 4M_c & 0 & \cdots \\
\cdots & \cdots & \cdots & \cdots & \cdots & \cdots & \cdots &
\cdots & \cdots & \cdots & \cdots
 \ea \right) .\eea   We then easily find the $\nu^c_n \nu^c_n$
component of the inverse which is just the inverse of each $2
\times 2$ matrix and it is zero.

For the most natural values of $M_D$ ($=\sqrt{\f{3}{4}} M_c$)
(Eqn. \ref{eq:mdbulk}) and for any value of $M_N$ satisfying $M_*
\ge M_N \ge M_c$, we get the effective right-handed Majorana mass
\bea M_{\rm eff} & = & \f{3\pi}{16} M_c. \eea    {\em Therefore in
this case, the effective Majorana mass for the right-handed
neutrino is ``naturally" given by the compactification scale $M_c
\sim 10^{14} \GeV$.}

%%%%%%%%%%%%%%%%%%%%%%%%%%%%%%%%%%%%%%%%%%%%%%%%%%%%%%%%%%%%%%%%%
\subsection{Brane Majorana Mass for $N$}
%%%%%%%%%%%%%%%%%%%%%%%%%%%%%%%%%%%%%%%%%%%%%%%%%%%%%%%%%%%%%%%%%

It is possible to imagine $M_N =0$. Suppose there is an additional
U(1) symmetry (lepton number) under which $N (N^c), \bar \chi^c$
carries a charge $1(-1), 3$. The bulk Majorana mass is forbidden.
Then if there is a brane localized field $S$, carrying U(1) charge
$-2$, we can write down the Majorana mass term on the
brane.\footnote{It is not clear whether it would be better to have
$S$ in the bulk or on the brane.  For definiteness here we
consider $S$ localized on the brane.} We consider the case of both
a brane localized and bulk field $\psi^c \supset \nu^c$.   As
before {\em we first consider $\nu^c$ localized on the brane}.

The most general superpotential has the mass terms given by  \bea
W & = & N \ \partial_5 N^c + \f{3 \pi^2}{M_*} \l_N \ S \ N \ N \
\delta (y) + 4 \pi \sqrt{\f{24 \pi^3}{M_*^3}} \l_D \ S \ \bar
\chi^c \ \psi^c \ N^c \ \delta (y-\f{\pi R}{2}), \eea which is
decomposed into KK modes as \bea W & = & \sum_{n=0}^{\infty}
\left[ (2n+1) M_c N_n N^c_n +  \sum_{m =0}^{\infty} \left( M_N N_n
N_m \right) + (-1)^n M_D \nu^c N^c_n \right] \eea with \bea M_N &
= & \f{3 \pi \la S \ra}{M_*} \l_N \ M_c \label{eq:mnbrane} \eea
and \bea M_D & = &  \f{\sqrt{24 \pi^2} \la S \ra}{M_*} \f{4 \pi
\la \bar \chi^c \ra}{M_*} \sqrt{\f{M_*}{M_c}} \l_D M_c \approx
\f{\sqrt{24 \pi^2} \la S \ra}{M_*} \sqrt{\f{M_*}{M_c}} M_c
\label{eq:mdsbrane} \eea where the last term is the ``natural"
value (see for comparison Eqn. \ref{eq:mdbrane}).

We now define \bea {\cal N} & = & (\nu^c, N_0, N^c_0, N_1, N^c_1,
\cdots,N_n,N^c_n,\cdots). \eea Then \bea W & = & \f{1}{2} \ {\cal
N}^T \ \hat{\cal M} \ {\cal N} \eea with \bea \hat{\cal M} & = &
\left( \ba{ccccccccc} 0 & 0 & M_D & 0 & -M_D &
\cdots & (-1)^n M_D & 0 & \cdots \\
0 & 2 M_N & M_c & 2 M_N & 0 & \cdots & 2M_N & 0 & \cdots
\\ M_D & M_c & 0 & 0 & 0 & \cdots & 0 & 0 & \cdots \\
0 & 2 M_N & 0 & 2 M_N & 3M_c & \cdots & 2M_N & 0 & \cdots \\
-M_D & 0 & 0 & 3M_c & 0 & \cdots & 0 & 0 & \cdots \\
\cdots & \cdots & \cdots & \cdots & \cdots & \cdots & \cdots &
\cdots & \cdots \\
(-1)^n M_D & 2M_N & 0 & 2M_N & 0 & \cdots & 2M_N & (2n+1)M_c &
\cdots \\
0 & 0 & 0 & 0 & 0 & \cdots & (2n+1)M_c & 0 & \cdots \\
\cdots & \cdots & \cdots & \cdots & \cdots & \cdots & \cdots &
\cdots & \cdots \ea \right). \eea

The effective right-handed neutrino mass is obtained similarly by
finding $({\hat{\cal M}}^{-1})_{\nu^c\nu^c}$ where  \bea
({\hat{\cal M}}^{-1})_{\nu^c\nu^c} & = & \frac{{\rm Det} {\cal
M}}{{\rm Det} \hat{\cal M}}. \eea   We obtain
 \bea {\rm Det} {\cal M} & = & \prod_{n=0}^{\infty} (-((2n+1)M_c)^2) \eea
and  \bea {\rm Det} \hat{\cal M} & = & -2 M_N  M_D^2 \left[
\sum_{n=0}^{\infty} \f{(-1)^n}{(2n+1)M_c} \right]^2 {\rm Det} {\cal M} \nn \\
& = & -\f{\pi^2 M_N  M_D^2}{8M_c^2} {\rm Det} {\cal M}. \eea
Therefore we obtain the effective Majorana mass for the
right-handed neutrino given by the expression \bea M_{\rm eff} & =
& \f{1}{|{(\hat{\cal M}}^{-1})_{\nu^c \nu^c}|} = \f{\pi^2 M_N
M_D^2}{8M_c^2}. \eea Note, this result is the same as in the case
of the bulk Majorana mass, when $M_N \ll M_c$ (see Eqn.
\ref{eq:meffmnsmall}). Finally, we have $M_N \sim \f{3 \pi \la S
\ra}{M_*} \ M_c$ (Eqn \ref{eq:mnbrane}). Thus $M_N$ is
``naturally" ${\cal O}(\f{3}{4} M_c)$ for $\f{4 \pi \la S
\ra}{M_*} \sim 1$.

We now obtain the following results for the effective right-handed
neutrino mass. When {\em $\nu^c$ is localized on the brane}, we
have $M_D \sim \f{\sqrt{24 \pi^2} \la S \ra}{M_*}
\sqrt{\f{M_*}{M_c}} M_c$ (Eqn. \ref{eq:mdsbrane}). Hence we can
get $M_{\rm eff} \sim M_c$, but only if we take $4 \pi \la S
\ra/M_* \sim 1/10$. However if we now take {\em $\nu^c$ in the
bulk}, using the ``natural" values for $M_D \sim \sqrt{\f{3}{4}}
M_c$ (see Eqn. \ref{eq:mdbulk}) and $M_N \sim \f{3}{4} M_c$, we
get $M_{\rm eff} \sim \f{9 \pi^2}{128} M_c$. In both cases, we
find results very similar to the case of bulk Majorana mass.  In
summary, {\em for $\nu^c$ localized on the brane an effective
right-handed neutrino mass of order $M_c$ can be accommodated,
while for $\nu^c$ in the bulk we ``naturally" obtain $M_{\rm eff}
\sim M_c$.}

%%%%%%%%%%%%%%%%%%%%%%%%%%%%%%%%%%%%%%%%%%%%%%%%%%%%%%%%%
\subsection{Left-Handed Tau Neutrino Mass}
%%%%%%%%%%%%%%%%%%%%%%%%%%%%%%%%%%%%%%%%%%%%%%%%%%%%%%%%%
Let us now complete the picture.   As discussed above, the
effective Majorana mass for $\nu^c$ (or its zero mode $\nu^c_0$ if
it is a bulk field) is given by $M_{\rm eff} \equiv \g M_c$ where
the parameter $\g =$ ${\cal O}(1)$ depends on the details of the
model. Note, only the zero mode of $\nu^c$ is effective in the
See-Saw mechanism.  We now construct the effective $2 \times 2$
mass matrix for $(\nu,\ \nu^c_0)$.   We have
\bea {\cal M}^{\rm eff}_{2 2} & = & \left( \ba{cc} 0 & \; m\\
 m & \; \g M_c \ea \right) \eea with $m = m_t$ at the GUT
scale.

We now consider the most general possibilities for having
$\psi^c$($\supset \nu^c$) in the bulk or on the PS brane.
Similarly, $\psi$($\supset \nu$) can be either in the bulk or on
the PS brane.  Note, gauge coupling unification can be affected by
the splitting of the left-handed $\psi$ and the right-handed
$\psi^c$ matter fields, if all KK modes are not in complete SO(10)
multiplets.

There are three possibilities for the third generation matter
fields configurations. In all cases, the condition for the
nontrivial fixed points is satisfied.
\begin{itemize}
\item $\psi$ and $\psi^c$ in the bulk:

Gauge coupling unification works fine and the See-Saw scale is
naturally of order $M_c$. However, there is a serious problem. The
natural value for the top quark Yukawa coupling has a volume
suppression with $\lambda = \l_t \sim \frac{(24 \pi^3)^{3/2}}{16
\pi^2} (\f{M_c}{M_*})^{3/2}$. Hence obtaining an order one
coupling is difficult.  At the very least, we would need to
localize the Higgs towards the PS brane. However this would
destroy the success of gauge coupling unification.   As a result
we are forced to place the third generation on the PS
brane.\footnote{Note, we could ameliorate the large suppression
factor $(\f{M_c}{M_*})^{\f{3}{2}}$ by decreasing the ratio
$M_*/M_c$.  In our previous paper \cite{Kim:2002im} we showed that
this ratio is fixed by the GUT threshold correction $\epsilon_3
\equiv (\alpha_3(M_{GUT}) - \tilde \alpha_{GUT})/\tilde
\alpha_{GUT} \approx -0.04$  required to fit the precision
electroweak data in the equivalent 4D theory [where
$\alpha_1(M_{GUT}) = \alpha_2(M_{GUT}) = \tilde \alpha_{GUT}$]. We
also note that the precise value of $\epsilon_3$ depends on the
squark, slepton and gaugino spectrum through electroweak threshold
corrections. It has recently been noted by Dermisek
\cite{Dermisek:susy2003} that as the universal GUT value of the
squark and slepton mass - $m_{16}$ - increases, with $m_{16} \gg
M_{1/2}$, the value of $\epsilon_3$ approaches zero. Moreover
using the results of Ref. \cite{Kim:2002im} we see that $M_*/M_c
\approx \exp(7 |\epsilon_3|/0.04)$ with $M_c \approx M_{GUT}
\exp(-5 |\epsilon_3|/0.04)$.  Hence as $|\epsilon_3|$ decreases
the compactification scale increases and the ratio $M_*/M_c$
decreases.  For example, with $\epsilon_3 = -0.02$ we have
$M_*/M_c \approx 33$ and $M_c \approx 2.5 \times 10^{15}$ GeV. In
this case we could put the third generation in the bulk, without
suffering any volume suppression in the Yukawa coupling, with the
See-Saw scale determined naturally by $M_c$ which is moderately
larger than before.  Thus whether the third generation should be a
bulk field or a brane field is somewhat model dependent.
\label{footnote:ftnote}} The first two families however can be
bulk fields with a welcome suppression of their Higgs Yukawa
coupling.

\item $\psi^c$ in the bulk and $\psi$ on the PS brane:

Again, the See-Saw scale is naturally of order $M_c$, however now
gauge couplings get a threshold correction which makes the
prediction worse.  The anomaly generated by massive KK modes of
$\psi^c$ is cancelled by a 5D Chern-Simons term \cite{Kim:2001at}.

\item $\psi$ in the bulk and $\psi^c$ on the PS brane:

The gauge coupling unification goes in the right direction. We get
a precise unification with a smaller ratio of $M_*/M_c$. The
effective right-handed neutrino mass needs an additional
suppression by $\f{4 \pi \la S \ra}{M_*}$ of order $\f{1}{10}$ to
give a See-Saw scale at $M_c$. Again there is an anomaly inflow
with a 5D Chern-Simons term.

\item $\psi$ and $\psi^c$ on the PS brane:

Gauge coupling unification works fine.  However, as above, the
See-Saw scale needs a suppression from $\f{4 \pi \la S \ra}{M_*}$
of order $\f{1}{10}$.

\end{itemize}

With the above caveats it is possible to obtain a light tau
neutrino mass $m_{\nu_\tau} \sim m^2/M_c$.  Each one of the above
choices, however, makes specific constraints on 5D SO(10) model
building.

%%%%%%%%%%%%%%%%%%%%%%%%%%%%%%%%%%%%%%%%%%%%%%%%%%%%%%%%%
\section{Conclusion}
%%%%%%%%%%%%%%%%%%%%%%%%%%%%%%%%%%%%%%%%%%%%%%%%%%%%%%%%%

There are now several physics issues related to the
compactification scale $M_c \approx 10^{14} \ \GeV$.

\begin{itemize}
\item Gauge coupling unification

In 4D, SUSY GUTs require a threshold correction $\epsilon \equiv (
\f{\alpha_3 (M_{GUT}) - \tilde{\alpha}_{GUT}}{
\tilde{\alpha}_{GUT}} ) \simeq -0.04$ in order to achieve a
perfect gauge coupling unification. In minimal SU(5), this results
from a color triplet Higgs mass $m_{H_3} \sim 10^{14} \ \GeV$. The
same result would be necessary in any SUSY GUT with an assumption
that there is no other threshold correction from the GUT breaking
sector.

In 5D, Orbifold SUSY GUTs determine the threshold correction as a
function of $M_c$ and $M_*$. There are no other free parameters.
Now the threshold correction requires the Higgs to be in the bulk
and determines the compactification scale $M_c \sim 10^{14} \
\GeV$.  In this framework, the 4D GUT scale is fictitious, however
it is related to the compactification and cutoff scales by the
approximate expression $M_{GUT} \approx (\f{M_*}{M_c})^{2/3} M_c$.
This relation determines the cutoff scale $M_* \sim 10^{17} \
\GeV$.

\item Neutrino mass

In 4D, the natural size of the right-handed tau neutrino Majorana
mass is determined from the higher dimensional operator
$\f{1}{M_{Pl}} \bf \overline{16} \ 16 \ \overline{16} \ 16$
 (or similarly $\f{1}{M_{Pl}} \bf \bar \chi^c \ \psi^c \ \bar \chi^c \ \psi^c$ in PS).  By
 replacing $\la \overline{16} \ra \sim M_{GUT}$ one obtains $M_{\nu^c} \sim
 \f{M_{GUT}^2}{M_{Pl}} \sim 10^{14} \ \GeV$.

In 5D, if $\nu^c$ is localized on the PS brane, we can accommodate
an effective right-handed neutrino mass of order $M_c$ either
using a U(1) symmetry with a symmetry breaking vev $\f{4 \pi \la S
\ra}{M_*} \sim 1/10$ to suppress the natural scale for the heavy
Dirac neutrino mass\footnote{This is the Froggatt-Nielsen
mechanism~\cite{Froggatt:1978nt} applied to neutrino masses.} or
with a small Majorana neutrino mass with $M_N/M_c \ll 1$ (or a bit
of both).

On the other hand, for $\nu^c$ in the bulk, the volume suppression
naturally gives $M_{\rm eff} \sim M_c \sim 10^{14} \ \GeV$ even
for $ \la \bar \chi^c \ra \sim M_*$.  Unfortunately, the same
volume suppression would naturally give a top Yukawa coupling
$\lambda_t \sim \frac{(24 \pi^3)^{3/2}}{16 \pi^2}
(\f{M_c}{M_*})^{3/2} \ll 1$.  This is a serious problem which can
be ameliorated by decreasing the volume factor
$(\f{M_*}{M_c})^{3/2}$.   One particular solution is testable at
the electroweak scale.  As discussed in Footnote
\ref{footnote:ftnote}, changing the universal squark and slepton
mass $m_{16}$ directly affects gauge coupling threshold
corrections at the electroweak scale and indirectly at the GUT
scale.  Moreover,  increasing $m_{16}$ has the effect of
increasing the compactification scale $M_c$,  decreasing the
cutoff scale $M_*$, hence decreasing the ratio $(M_*/M_c)^{3/2}$.
It is encouraging that the scale necessary for the atmospheric
neutrino oscillations (which is the tau neutrino mass in SO(10))
can be the same as the compactification scale $M_c \sim 10^{14 -
15}$ GeV, necessary for a successful unification of gauge
couplings.

Finally, it is interesting to note that part of this result could
simply have been obtained using an effective higher dimensional
right-handed neutrino mass operator on the PS brane given by \bea
W & = & \f{C}{2 M_{*}^n} \bf \bar \chi^c \ \psi^c \ \bar \chi^c \
\psi^c \; \delta(y - \f{\pi R}{2}) \eea with $n$ = 1, $C = 16
\pi^2 c$ (for $\psi^c$ on the PS brane) or $n$ = 2, $C = 24 \pi^3
c$ (for $\psi^c$ in the bulk).   Thus when $4 \pi \la \bar \chi^c
\ra \sim M_*$  in the right-handed neutrino direction and $c \sim
1$ (i.e. the ``natural" values) we find a right-handed neutrino
mass \bea W & = & \f{1}{2} M_{\rm eff} \ \nu^c \ \nu^c \eea  with
\bea M_{\rm eff} & \sim & \left(
\begin{array}{c} M_* \\ \f{3}{4} M_c \end{array} \right) \;\; {\rm for} \;\; \psi^c
\;\; \left( \begin{array}{c}    {\rm on \; the \; PS \; brane}  \\
\;\; {\rm in \; the \; bulk} \end{array} \right). \eea   Once
again we find that the value of $M_{\rm eff}$ for $\psi^c$ in the
bulk is of order $M_c$, but for $\psi^c$ on the PS brane we can
only obtain this desired value with a small value of $c \sim
10^{-3}$. However note, with this effective operator analysis we
cannot realize the possibility of suppressing $M_{\rm eff}$ (for
$\psi^c$ on the PS brane) with a small Majorana mass.

\item 5D Planck scale

There is another very interesting relation between the 5D Planck
scale $M_{\rm 5D}$ and the cutoff scale $M_*$. In the presence of
the extra dimension, the 4D Planck scale is a derived scale which
is determined by the 5D Planck scale $M_{\rm 5D}$ and the
compactification scale $M_c$. The relation is $M_{\rm Pl}^2 =
\f{M_{\rm 5D}^3}{M_c}$.  Now for $M_c \sim 10^{14} \ \GeV$, we get
$M_{\rm 5D} \sim 10^{17} \ \GeV$.   Note,  this is the same as the
cutoff scale $M_*$ determined assuming gauge coupling unification.

\end{itemize}

In conclusion, in this paper we have discussed several different
possible frameworks for obtaining a See-Saw mechanism in 5D SUSY
SO(10).   If the right-handed tau neutrino is in the bulk then we
find that the ``natural" value of the See-Saw scale is given by
the compactification scale $M_c \sim 10^{14}$ GeV.  As noted in
the text, this scenario has a serious problem with a very small
top quark Yukawa coupling.  This problem can be overcome, however,
with the bottom line effect of moderately increasing $M_c$ to
about $10^{15}$ GeV. On the other hand, if the right-handed tau
neutrino is on the Pati-Salam brane, then the ``natural" value for
the See-Saw scale is greater than the compactification scale.
However, using a Froggatt-Nielsen like mechanism and/or a small
Majorana neutrino mass $M_N \ll 1$, a See-Saw scale of order $M_c$
can be obtained.

Note, the values of the compactification scale $M_c$ and the
cutoff scale $M_*$ are determined by gauge coupling unification
\cite{Kim:2002im}.   In addition, we note that the cutoff scale is
the same as the 5D Planck scale determined using the observed the
4D Planck scale and the compactification scale. This triple
significance for a scale around $10^{14} \ \GeV$ is either an
amazing coincidence or very profound.

Finally, we have only discussed the effective right-handed
Majorana mass appropriate for one family of quarks and leptons in
this paper. Further discussion of neutrino mixing angles within
this framework must wait until a three family model is
constructed.

%%%%%%%%%%%%%%%%%%%%%%%%%%%%%%%%%%%%%%%%%%%%%%%%%%%%%%%%%%%%%%%%%%%%%%%
\acknowledgments   Partial support for this work came from DOE
grant\# DOE/ER/01545-842.  S.R. would also like to acknowledge the
support from the Institute for Advanced Study through a grant in
aid from the Monell Foundation.
%%%%%%%%%%%%%%%%%%%%%%%%%%%%%%%%%%%%%%%%%%%%%%%%%%%%%%%%%%%%%%%%%%%%%%%

%%%%%%%%%%%%%%%%%%%%%%%%%%%%%%%%%%%%%%%%%%%%%%%%%%%%%%%%%
\appendix
%%%%%%%%%%%%%%%%%%%%%%%%%%%%%%%%%%%%%%%%%%%%%%%%%%%%%%%%%

%%%%%%%%%%%%%%%%%%%%%%%%%%%%%%%%%%%%%%%%%%%%%%%%%%%%%%%%%
\section{5D Fermion Mass Operator}
%%%%%%%%%%%%%%%%%%%%%%%%%%%%%%%%%%%%%%%%%%%%%%%%%%%%%%%%%

%%%%%%%%%%%%%%%%%%%%%%%%%%%%%%%%%%%%%%%%%%%%%%%%%%%%%%%%%
\subsection{5D Mass in terms of 4D Weyl spinors}
%%%%%%%%%%%%%%%%%%%%%%%%%%%%%%%%%%%%%%%%%%%%%%%%%%%%%%%%%

To express 5D mass terms in terms of 4D Weyl spinor, we summarize
our spinor convention and charge conjugation (especially for the
5D Majorana mass).

Dirac $\g$ matrix in 5D is \bea \g_0 & = & \left( \ba{cc} 0 & 1 \\
1 & 0  \ea \right), \ \ \ \g_i = \left( \ba{cc} 0 & -\sigma_i \\
\sigma_i & 0 \ea  \right), \ \ \
\hat{\g_5} = \left( \ba{cc} -i & 0 \\
0 & i \ea \right),
 \eea where $\sigma_i$ with $i=1,2,3$
is the $2 \times 2$ Pauli matrix. Each element is thus a $2 \times
2$ matrix. Note that $\hat{\g_5} = i \g_5$ for the usual $\g_5$
defined in 4D theory. Now $\g_M = \{ \g_{\mu}, \hat{\g_5} \}$ with
$M=0,1,2,3,5$ forms a Clifford algebra in 5D \bea \{ \g_M, \g_N \}
& = & 2 g_{MN}. \eea

Generators of Lorentz transformations for 5D spinors are given by
\bea \Sigma_{MN} & \propto & i  [ \g_M, \g_N ]. \eea

A 5D spinor and its complex conjugate is given by \bea \psi & = & \left( \ba{c} \chi_1 \\
i \sigma_2 \chi_2^* \ea \right), \ \ \
\psi^*  =  \left( \ba{c} \chi_1^* \\
i \sigma_2 \chi_2 \ea \right). \eea

Let us consider charge conjugation. The Dirac $\g$ matrices have
the property that $\g_M^* = \g_M$ for $M=0,1,3$ and $\g_M^* =
-\g_M$ for $M=2,5$. For $C = \g_2 \hat{\g_5}$, we have \bea C
\g_M^* C^{-1} = \g_M, \eea with $C^{-1} = - C$.

Charge conjugation is defined as \bea \psi_c & \equiv & C \psi^* = \left( \ba{c} \chi_2 \\
-i \sigma_2 \chi_1^* \ea \right). \eea

It is easy to check that $\psi_c$ transforms as the same as $\psi$
under the 5D Lorentz transformation.

Using the property $\g_0 \g_M^{\dagger} \g_0 = \g_M$, we get \bea
\bar{\psi} \psi & = & ( \chi_1^{\dagger} \ \ - \chi_2^{T} i
\sigma_2 ) \left( \ba{cc} 0 & 1 \\ 1 & 0 \ea \right) \left( \ba{c}
\chi_1 \\ i \sigma_2 \chi_2^* \ea \right) \nn \\
& = & \chi_2^{T} (-i\sigma_2) \chi_1 + {\rm h. c.} \nn \\
&& \rightarrow  \chi_2 \chi_1
+ {\rm h.c.}, \\
\bar{\psi} \psi_c & = & ( \chi_1^{\dagger} \ \ - \chi_2^{T} i
\sigma_2 ) \left( \ba{cc} 0 & 1 \\ 1 & 0 \ea \right) \left( \ba{c}
\chi_2 \\ -i \sigma_2 \chi_1^* \ea \right) \nn \\
& = & \chi_2^{T} (-i\sigma_2) \chi_2 + \chi_1^{\dagger}
(-i\sigma_2) \chi_1^* \nn \\
&& \rightarrow  \chi_2 \chi_2 - \chi_1 \chi_1 + {\rm h.c.}. \eea

We use $\chi_1 \rightarrow N$ and $\chi_2 \rightarrow N^c$. The
corresponding Dirac mass in a supersymmetric theory is given by
\bea W & = & M_D N N^c, \eea and the Majorana mass is \bea W & = &
\f{1}{2} M_N ( - N N + N^c N^c ). \eea

%%%%%%%%%%%%%%%%%%%%%%%%%%%%%%%%%%%%%%%%%%%%%%%%%%%%%%%%%
\subsection{Equivalence of 5D Dirac Mass and Majorana Mass}
%%%%%%%%%%%%%%%%%%%%%%%%%%%%%%%%%%%%%%%%%%%%%%%%%%%%%%%%%

Since a 5D Majorana mass term requires two 4D Weyl spinors, it is
natural to ask whether there is a transformation which changes a
5D Dirac Mass to a 5D Majorana Mass, and vice versa.

Let us start from the theory with one hypermultiplet composed of
two 4D chiral multiplets $N_+$ and $N_-$ with a 5D Dirac mass.
Neglecting the boundaries, in 5D, we have \bea W & = & N_+
\partial_5 N_- + m N_+ N_-, \eea which is the same as \bea W & = &
\f{1}{2} (N_+ \partial_5 N_- - N_-
\partial_5 N_+ ) + m N_+ N_-, \eea
up to a total derivative.  We can re-define the fields,
\bea N & = & \f{1}{\sqrt{2}} ( N_+ - N_- ), \nn \\
N^c & = & \f{1}{\sqrt{2}} ( N_+ + N_- ). \eea  The 5D
superpotential is then expressed in terms of $N$ and $N^c$ by \bea
W & = &  W_{\rm bulk} + W_{\rm boundary} \\
W_{\rm bulk} & = & \f{1}{2} (N \partial_5 N^c - N^c
\partial_5 N ) + \f{1}{2} m (-N N +  N^c N^c) \nn \\
W_{\rm boundary} & = & \f{1}{4} \partial_5 ( - N N + N^c N^c ).
\nn \eea Hence, we obtain the Majorana mass from the Dirac mass by
the field re-definitions, except for the last term which is a
total derivative in 5D. Therefore, it is shown that the Dirac mass
and the Majorana mass are equivalent in 5D up to boundary terms.
Finally, in the presence of the boundaries, specifically for the
orbifold compactification, the Majorana mass term can be
re-expressed in terms of the Dirac mass term plus the residual
effects given by \bea W_{\rm boundary} & = & \f{1}{4}
( -N N + N^c N^c ) \left[-\delta (y) + \delta ( y-\f{\pi R}{2} )\right] \nn \\
& = & \f{1}{4} \left[ N N \delta (y) + N^c N^c \delta ( y-\f{\pi
R}{2} )\right], \eea for $N_{+-}$ and $N^c_{-+}$.

%%%%%%%%%%%%%%%%%%%%%%%%%%%%%%%%%%%%%%%%%%%%%%%%%%%%%%%%%
\subsection{Useful formulae}
%%%%%%%%%%%%%%%%%%%%%%%%%%%%%%%%%%%%%%%%%%%%%%%%%%%%%%%%%

\bea \sum_{n=-\infty}^{\infty} \f{1}{x+n} & = & \pi \cot (\pi x).
\eea

\bea \sum_{n=1}^{\infty} \left[ \f{1}{x+n} + \f{1}{x-n} \right] &
= & -\f{1}{x} + \pi \cot (\pi x). \eea

\bea \sum_{n=0}^{\infty} \f{x}{x^2 + (2n+1)^2} & = & \f{\pi}{4}
\tanh \left( \f{\pi x}{2} \right). \eea

\bea \sum_{n=0}^{\infty} \f{(-1)^n}{2n+1} & = & \f{\pi}{4}. \eea

%%%%%%%%%%%%%%%%%%%%%%%%%%%%%%%%%%%%%%%%%%%%%%%%%%%%%%%%%%%%%%%%%%%

%%%%%%%%%%%%%%%%%%%%%%%%%%%%%%%%%%%%%%%%%%%%%%%%%%%%%%%%%
\end{document}